# RCFT : Re-Clustering Formation Technique in Hierarchical Sensor Network

Boseung Kim
Department of Computing
Soongsil University
Seoul, South Korea

Joohyun Lee
Department of Computing
Soongsil University
Seoul, South Korea

Yongtae Shin
Department of Computing
Soongsil University
Seoul, South Korea

*Abstract—* Because of limited energy of nodes, an important issue for sensor network is efficient use of the energy. The clustering technique reduces energy consumption as cluster head sends sensed information to a sink node. Because of such character of clustering technique, electing cluster head is an important element for networks. This paper proposes RCFT(Re-Clustering Formation Technique) that reconstruct clusters in hierarchical sensor networks. RCFT is a protocol that reconstructed clusters considering position of a cluster head and nodes in randomly constructed clusters. And this paper demonstrated that clusters are composed evenly through simulation, accordingly this simulation shows the result reducing energy consumption.

*Keywords-Wireless Sensor Networks, Clustering*

## I. INTRODUCTION

AS the interest in the surroundings of Ubiquitous increases recently, we, also, pay much attention to the sensor-network, which composes one of the components in Ubiquitous. Sensor-nodes, which have the limited energy, are mostly set in the area where is dangerous or not easily accessible[1].

Accordingly, as it is very difficult for sensor-nodes to be replaced even after the energies of them are used up once set in, it is the most important part for study in the field to prolong the life-span of the sensor network through the proficient use of energy.

Considering the trait that gathering of data is required in order to decrease the waste of energy caused by the double transmittance of information between the adjacent sensor-nodes, The Routing Protocol based upon cluster has much of advantages.

Selecting the head of cluster is essential in the hierarchical Routing Protocol based upon cluster, so the proper selecting the head enables us to save the electrical power as well as to disperse the waste of energy.

This paper suggests the devices to select differently the head taking the positions of the heads of cluster and the distance between sensor-nodes. The suggested technique is to select the heads of cluster optionally. And then the heads of cluster applied with the techniques are diversified in order.

The cluster settled in this way is not to be re-organized at every round, but to be fixed by the end of its life-span. This research aims for properly dividing the range of cluster, decreasing the waste of energy by fixing the cluster, and prolonging the life-span of the sensor network.

## II. RELATED STUDY

### A. LEACH(Low-Energy Adaptive Clustering Hierarchy)

LEACH[2] is the technique of Routing based upon clustering for the purpose of dispersing the loads of energy between the sensor-nodes. In LEACH, the sensor-nodes are being composed by themselves, and one sensor-node plays a part of head.

In case of functioning as head of cluster, the sensor nodes waste energy much more than the ordinary senor nodes because they should collect and summarize data from other sensor nodes, and transmit it to BS. So, assuming that all the sensor nodes have the identical level of energy, the sensor nodes selected as the cluster heads exhaust out fast.

Therefore, LEACH makes many sensor-nodes within the cluster take the position of the cluster heads by shift to prevent this situation. Also, LEACH exercises the regional absorption of data in order to absorb the data from cluster to BS, which helps to save the energy and to make life-span of the system longer.

LEACH is composed of rounds, and each round has two(2) stages; 'set-up' stage, in which cluster get organized, and 'steady-state' stage, in which many TDMA frames get formed. LEACH is the basic technique for the hierarchical sensor network. So far LEACH-C[3], TEEN[4], APTEEN[5], which are gotten rid of weak points of LEACH, have been introduced.

### B. LEACH-C

LEACH-C(LEACH-Centralized) is also the technique of Routing Protocol based upon clustering. Though it is similar with LEACH, LEACH-C is the method that synch selects the cluster heads according to the information on sensor-nodes' position and the holding amount of energy with regard to selecting the cluster heads.

During the stage of 'set-up', each sensor-node transmits the information on its present position and the level of energy to BS. On receiving the message, BS calculates the average values of energy level of all the sensor-nodes, and then decides the heads of cluster by minimizing the total sum of the distances between the cluster heads and non cluster heads.

When cluster is established, BS broadcasts messages including the ID of cluster heads to each sensor-node, and It is the sensor-nodes having the identical ID to the ID in the message that are to be selected as cluster heads.

The strong point of LEACH-C is that it can leads in the equal waste of energy between sensor-nodes by inducing the cluster heads into the centre of cluster.

However, each sensor-node should recognize the information on its position, for which each sensor-node should be loaded with GPS receiver set. This apparatus will make the price of sensor-nodes increase highly. As quantity of sensor-nodes to be needed for the network ranges from hundreds to hundred-thousands, increase in the price of sensor-nodes is not appropriate[6].

## III. CLUSTERING ALGORITHM SUGGESTED

### A. Problem of the established Clustering Algorithm

LEACH re-organizes cluster at every termination of one round, in this process the cluster heads is selected at random except for the sensor-nodes to have been already selected as heads. Accordingly, cluster can be divided equally, or not as shown in figure1)[7].

If cluster is not divided properly, each sensor-node' waste level of energy would increase as well as not be in order.

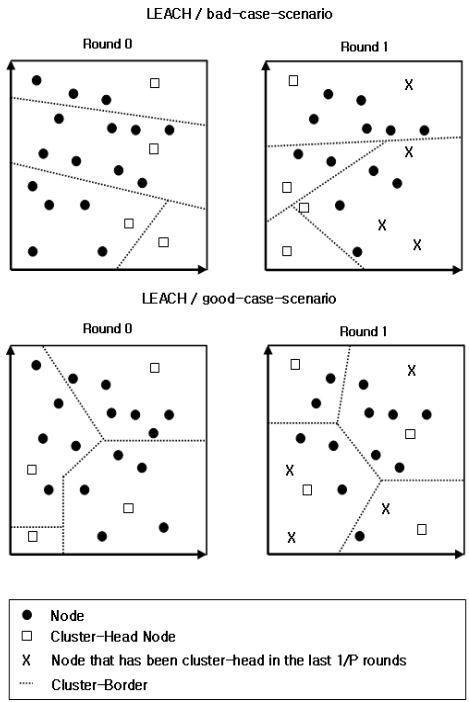

Figure 1. Division of cluster in LEACH

### B. RCFT(Re-Clustering Formation Technique)

LEACH re-organizes cluster at every termination of one round, in this process the cluster heads is selected at random except for the sensor-nodes to have been already selected as heads. Accordingly, cluster can be divided equally, or not as shown in figure1)[7].

If cluster is not divided properly, each sensor-node' waste level of energy would increase as well as not be in order.

*1) Abstract of RCFT:* RCFT suggested in the paper selects randomly cluster heads at first, then re-selects cluster heads considering the numbers of hops between each cluster heads , and the numbers of hops of cluster nodes farthest away from the cluster heads. After selecting cluster heads, RCFT re-organizes cluster which is to be fixed till the termination of network's life-span.

*2) Operation of RCFT:* After broadcasting the broadcast message, the sensor-nodes selected as the first cluster heads wait for response for a while, and when received response, they inspect whether the responses are identical with ones from the same sensor-nodes, which responded before.

If the response is the first time, they record the response of head with most small numbers of hops, and also record the information of the sensor-node with the most counting values among the responses on sensor-nodes. If there are over two of sensor-nodes having the most counting values, the information of sensor-node, which responded at the last, is to be recorded as it means that the sensor-node, which responded at the last, is the farthest away.

If the sensor-nodes selected as the first cluster heads receive responses from all the nodes, they subtract the numbers of hops of the farthest sensor-node from the numbers of hops of the closest head. If the calculation results in plus, they move for the direction of the closest head as many as the numbers of hops in the values; If the calculation results in minus, they move for the direction of the farthest sensor-node as many as the numbers of hops in the values. Given the ttl value as result value, and if ttl value makes 0, the sensor-nodes are to be selected as new head cluster. If the result value makes 0, the first cluster head does not move to become cluster head again.

(Figure2) shows the example of technique suggested.

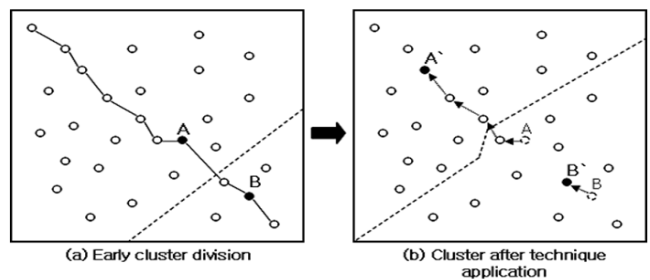

Figure 2. Example of RCFT

(a) in left side shows that A and B are selected as the first cluster head. The cluster of (a) is not in order, which is common when the cluster heads are selected at random. (b) in right side shows the condition of division of cluster after technique suggested applied. Head A moved 4 hops for the direction of nodes, and head B moved 1 hop for the direction of cluster. It is seen that the irregularly divided cluster(a) can be divided comparatively in order.

## IV. EVALUATION OF EFFICIENCY

In order to show the superiority of RCFT, it is needed to compare and analyze the average numbers of sensor-nodes which each one cluster in LEACH and in the technique suggested have respectively. Also it is to be done to calculate the average distance between LEACH and LEACH-C, and between the cluster head in the technique suggested and the sensor-nodes belonging in, and to compare and analyze the level of energy consumed by sensor-node while the rounds repeat.

### A. Condition of experiment

Table 1 shows the condition for evaluating the efficiency. Under the circumstance of 100m X 100m, the total numbers of sensor-nodes is 100units, and the cluster heads compose 5 units, which is 5% of the total sensor-nodes.

TABLE I. TABLE TYPE STYLES

| Classification | Factor | Set-up |
|---|---|---|
| **Work condition** | *Language* | *Visual C++* |
| | *OS* | *Windows XP Professional* |
| **Experiment condition** | *Rage of sensor-field* | *100m X 100m* |
| | *Total numbers of nodes* | *100 units* |
| | *Numbers of heads* | *5* |
| | *Position of BS* | *(50, 500)* |
| | *Times of experiment* | *20 Round X 10times* |
| | *size of packet* | *2000 bit* |

To analyze whether the cluster is divided in order, calculated are the distances between cluster head and the each node belonging, and the numbers of belonging nodes at every termination of each round. The average distance was calculated by dividing the total sum of the distances between cluster head and the each belonging nodes by the numbers of the belonging nodes, and this experiment was conducted 10 times based upon the criteria of 20 rounds.

The same numerical Formula used with LEACH was adopted for analyzing the consuming of LEACH, LEACH-C, and RCFT.

### B. Result of experiment and Analysis

(Figure 3) shows the average numbers of nodes per cluster. It can be estimated that the closer to 20 units are the average numbers of nodes per cluster, the cluster is divided more regularly. In (Figure 3), it can be found out that as RCFT comes closer to the average values, 20 than LEACH, the gaps get small.

(Figure 4) shows the distribution of the numbers of nodes per cluster. As the cluster heads of LEACH are selected randomly, the numbers of nodes belonging in cluster are irregular. Therefore, as shown in (Figure 4), various numbers of node result in. Especially, numbers of nodes under 10 and over 31 come out frequently. On the contrary, numbers of nodes ranging from 16 to 25, which can be considered as relatively good result, are found almost over 50% in the technique suggested.

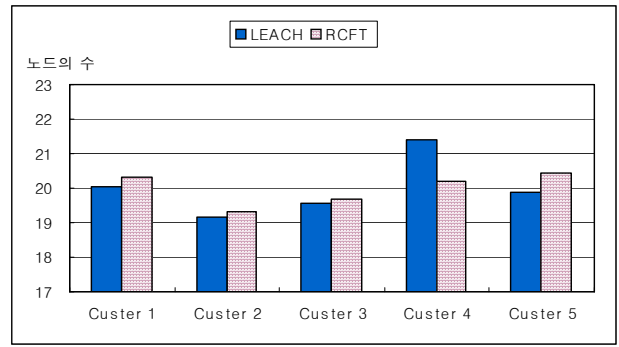

Figure 3. Distribution of the numbers of nodes per cluster

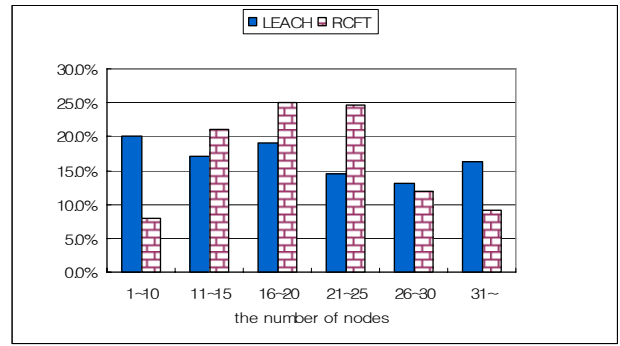

Figure 4. Distribution of the numbers of nodes per cluster

(Figure 5) shows the average distances between cluster heads and all the belonging node in LEACH, the technique suggested, and LEACH-C which uses separate information of position. As the longer is average distances, the wider cluster is, and the shorter is average distance, the closer cluster heads come to the center of cluster, it can be demonstrated that the more efficient does cluster get, the shorter is the average distance.

In case of LEACH, 21.11m was measured as the average distance. On the contrary, the average distance was 20.68m in LEACH-C using the information of position. The technique suggested had 20.88m as average, which is a little worse than LEACH-C, but still shows similar capability.

(Figure 6) shows the average energy-consuming quantity of nodes. LEACH-C using the positional information saved about 20% more of energy waste than LEACH. Even though at the first stage of 20 round RCFT caused almost two(2) times more of energy waste than other techniques as it organizes cluster once more than others at first, it shows gradually low rates of increase. After 120 round, the energy waste of RCFT became smaller than LEACH, and it increase in the similar ratio of LEACH-C.

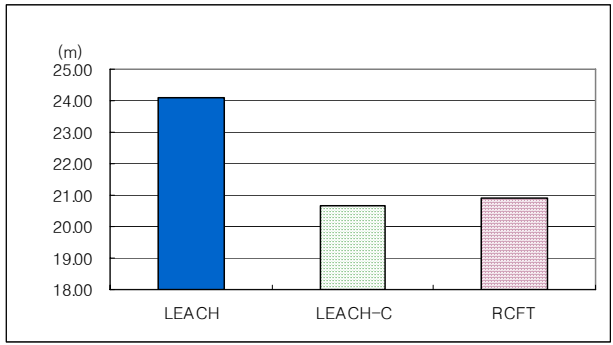

Figure 5. Average distance between clusters and nodes

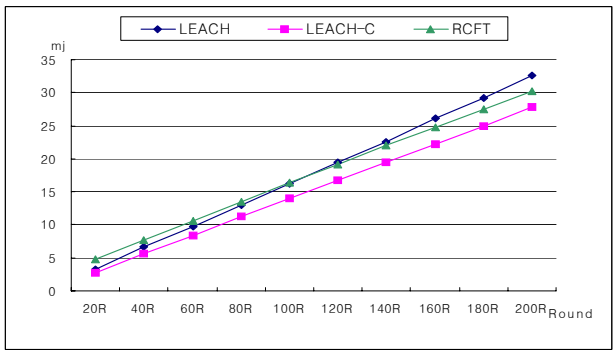

Figure 6. Average amount of energy-waste of nodes

### C. Analysis on experimental results

The result of experiment shows that the technique suggested is more efficient than LEACH with regard to division of cluster. The average numbers of nodes was closer to the average value in the technique suggested than LEACH, and the average distance between cluster heads and nodes became shorter in the technique suggested than LEACH.

In addition, the technique suggested enabled it to get the similar value with LEACH-C using the separate positional information. Even though at the first stage of 20 rounds RCFT caused more of energy waste than other techniques, it showed gradually low rates of increase than LEACH. After 120 rounds, the energy waste of RCFT became much smaller than LEACH.

## V. CONCLUSION

This paper suggests Re-clustering Formation Technique in the hierarchical sensor network. The technique suggested is to disperse and re-organize cluster heads considering the numbers of hops between the clusters organized randomly and the belonging nodes for the sake of the efficient division of clusters.

Network model was realized for analyzing the efficiency of the suggestion. The analysis on the efficiency shows that division of clusters is more efficient in the technique suggested than in the established techniques, which can save the waste of energy.

Also, It was shown that the technique suggested is not much different from the one using separate positional information. Through minimizing the energy-wastes of the entire network with the aid of the Re-clustering Formation Technique suggested in the paper, it is possible to accomplish more efficient surrounding of communications in the hierarchical sensor network.

### REFERENCES

[1] Ian F. Akyildiz, Weilian Su, Yogesh SanKarasubramaniam, and Erdal Cayirci, "A survey on Sensor Networks, "IEEE Communications Magazine, vol.40, No.8, pp.102-114, August 2002.

[2] Wendy Rabiner Heinzelman, Anantha Chandrakasan, and Hari Balakrishnan, "Energy-Efficient Communication Protocol for Wireless Microsensor Networks", Proceedings of the Hawaii International Conference on System Sciences, January 2000.

[3] Endi B. Heinzelman, Anantha P. Chandrakasan, and Hari Balakrishnan, "An Application-Specific Protocol Architecture for Wireless Microsensor Networks", IEEE Transactions On Woreless Communications, Vol. 1, No. 4, October 2002.

[4] Arati Manjeshwar, Dharma P. Agrawal, "TEEN: A Routing Protocol for Enhanced Efficiency in Wireless Sensor Networks," ipdps, p. 30189a, 15th International Parallel and Distributed Processing Symposium (IPDPS'01) Workshops, 2001.

[5] A. Manjeshwar and D.P. Agrawal, "APTEEN: A Hybrid Protocol for Efficient Routing and Comprehensive Information Retrieval in Wireless Sensor Networks," in the Proceeding of the 2nd International Workshop on Parallel and Distributed Computing Issues in Wireless Networks and Mobile Computing, Ft.Lauderdale, FL, April 2002.

[6] Mohammad Ilyas, Imad Mahgoub, "Handbook of Sensor Networks: Compact Wireless and Wired Sensing Systems", CRC PRESS, 01, 2006.

[7] M. J. Handy, M. Haase, D. Timmermann, "Low Energy Adaptive Clustering Hierarchy with Deterministic Cluster-Head Selection", IEEE, 2002.

#### AUTHORS PROFILE

B. Kim. Author is with the Department of Computing, Ph.D. course, Soongsil University, Seoul, Korea. His current research interests focus on the communications in wireless sensor networks (e-mail:bskim@cherry.ssu.ac.kr).

J. Lee. Author is with the Department of Computing, M.Sc. course, Soongsil University, Seoul, Korea. His current research interests focus on the communications in wireless sensor networks (e-mail:jhlee@cherry.ssu.ac.kr).

Y. Shin. Author was with the Computer Science Department M.Sc. and Ph.D., University of Iowa. He is now with the Professor, Department of Computing, Soongsil University. (e-mail: shin@ssu.ac.kr).